\documentclass[12pt,preprint]{aastex}

\begin{document}
\title{Hexadecapole Approximation in Planetary Microlensing}

\author{Andrew Gould}
\affil{Department of Astronomy, Ohio State University,
140 W.\ 18th Ave., Columbus, OH 43210, USA; 
gould@astronomy.ohio-state.edu}

\begin{abstract}
The frequency of microlensing planet detections, particularly
in difficult-to-model high-magnification events, is increasing.
Their analysis can require tens of thousands of processor hours
or more, primarily because of the high density and high precision
of measurements whose modeling requires time-consuming finite-source 
calculations.  I show that a large fraction of these measurements,
those that lie at least one source diameter from a caustic or the
extension from a cusp,
can be modeled using a very simple hexadecapole approximation,
which is one to several orders of magnitude faster than full-fledged
finite-source calculations.  Moreover,
by restricting the regions that actually require finite-source
calculations to a few isolated ``caustic features'',
the hexadecapole approximation will, for the first time, 
permit the powerful ``magnification-map''
approach to be applied to events for which the planet's orbital motion
is important.

\end{abstract}


\section{{Introduction}
\label{sec:intro}}

Microlensing planets are being discovered at an accelerating rate,
with one being reported in 2003 \citep{ob03235}, 
three in 2005 \citep{ob05071,ob05390,ob05169}, two in 2006
\citep{gaudi08}, and perhaps as
many as six in 2007.  It has been a huge challenge for modelers to
keep up with these discoveries, in large part because the computing
requirements are often daunting: the parameter space is large, the
$\chi^2$ surface is complex and generally contains multiple minima,
and the magnification calculations are computationally intensive.  
Proper treatment of individual events can require tens of 
thousands processor hours, or more.  Indeed, some potential planetary 
events have still not been fully modeled because of computational 
challenges.

Planetary microlensing events are recognized through two
broad channels, one in which
the lightcurve perturbation is generated by the so-called ``planetary caustic''
that is directly associated with the planet \citep{gouldloeb92} and the
other in which it is generated by the ``central
caustic'' that is associated with the host star \citep{griestsafi}.
The former events are relatively easy to analyze and indeed the event
parameters can be estimated reasonably well by inspection of the lightcurve.
The latter events are generally much more difficult. 

There are several interrelated reasons for this.  First, planets anywhere
in the system can perturb the central caustic.  This is why these events are
avidly monitored, but by the same token it is often not obvious
without an exhaustive search which planetary geometry 
or geometries are responsible for the perturbation.
Second, this very fact implies that several members of a multi-planet
system can be detected in central-caustic events \citep{gaudi98}.
Multiple planets create a larger, more complicated parameter space,
which can increase the computation time by a large factor.
Even if there are no obvious perturbations caused by a second planet,
an exhaustive search should be conducted to at least place upper limits
on their presence.  Third, if the source probes the central caustic, it
is ipso facto highly magnified.  Such events are brighter and more 
intensively monitored than typical events and so have more and higher-quality
data.  While such excellent data are of course a boon to planet searches,
they also require more and more-accurate computations, which requires
more computing time.  Fourth, central-caustic planetary events are basically
detectable in proportion to the size of the caustic, which roughly
scales $\propto q/|b-1|$, where $q$ is the planet/star mass ratio
and $b$ is the planet-star separation in units of the Einstein ring.
Thus, these events are heavily biased toward planets with characteristics
that make the caustic big.  Such big caustics can undergo subtle changes
as the planet orbits its host, and in principle these effects can be
measured, thus constraining the planet's properties.  Exploration of these
subtle variations requires substantial additional computing time.
Finally, there is also a bias toward long events, simply because
these unfold more slowly and so increase the chance that they will
be recognized in time to monitor them intensively over the peak.
Such long events often display lightcurve distortions in their wings
due to the Earth's orbital motion which, if measured, can further
constrain the planet properties.  However, probing this effect (called
``microlens parallax'') requires yet another expansion of parameter space.
Moreover, the ``parallax'' signal must be distinguished from ``xallarap''
(effects of the source orbiting a companion), whose description
requires a yet larger expansion of parameter space.

There are two broad classes of binary-lens (or triple-lens)
magnification calculations: point-source and finite-source.  The
former can be used whenever the magnification is essentially constant
(or more precisely, well-characterized by a linear gradient) over
face of the source, while the latter must be used when this condition
fails.  Point-source magnifications can be derived from the solution
of a 5th (or 10th) order complex polynomial equation and are
computationally very fast.  When the point-source is well separated
from the caustics (as it must be to satisfy the linear-gradient condition)
then this calculation is also extremely robust and accurate.  

The main computational challenge in modeling planetary events comes
from the finite-source calculations.  Almost all integration schemes
use inverse-ray shooting, which avoids all the pathologies of the caustics
by performing an integration over the {\it image} plane (where the
rays behave smoothly) and simply asks which of the rays land on the
source.  The problem is that a large number of rays must be ``shot''
to obtain an accurate estimate of the magnification, which implies that
high-quality data demand proportionately longer computations.  Of course,
the higher the magnification, the larger the images, and so the more
rays are required.  There are various schemes to expedite inverse
ray-shooting, including clever algorithms for identifying the regions
that must be ``shot'' and pixelation of the source plane.  The
bottom line is, however, that the overwhelming majority of computation
time is spent on finite-source calculations.

Here I present a third class of binary-lens (or triple-lens) computation
that is intermediate between the two classes just described:
the hexadecapole approximation.  \citet{pejcha07} expand 
finite-source magnification to hexadecapole order and illustrate that
the quadrupole term by itself can give quite satisfactory numerical results.
In this paper, I develop a simple prescription for evaluating this
expansion.  While this algorithm
is about 10 times slower than point-source
calculations, it is one to several orders of magnitude faster than
finite-source calculations.  The method can be applied whenever
the source center is at least two source radii from a caustic
or the extension from a cusp.
Typically, well over half the non-point-source points satisfy this
condition, meaning that the method can reduce computation times by
a factor of several.  Moreover, by isolating the small regions of the 
lightcurve where finite-source calculations must be used, the method opens 
up the possibility that the finite-source computations themselves 
can be radically expedited for the special, but very interesting, class
of events in which planetary orbital motion is measured.

\section{Hexadecapole Approximation}

If the source does not straddle a caustic, then the magnification field 
can be Taylor expanded around the source center $(x_0,y_0)$ as a function of
coordinate position $(x,y)$ (all distances being expressed in units
of the Einstein radius),
\begin{equation}
A(x,y) = \sum_{n=0}^\infty\sum_{i=0}^n A_{ni} (x-x_0)^i (y-y_0)^{n-i}.
\label{eqn:axy}
\end{equation}
We wish to evaluate $A_{\rm finite}(\rho)$, the magnification of a 
source of radius $\rho$,
\begin{equation}
A_{\rm finite}(\rho,x_0,y_0) 
\equiv {\int_0^\rho dw\,\int_0^{2\pi} d\eta \,A(w,\eta)\over \pi\rho^2},
\label{eqn:afinite0}
\end{equation}
in terms of the $A_{ni}$.  Here $(w,\eta)$ are polar coordinates:
$(w\cos\eta,w\sin\eta) \equiv (x-x_0,y-y_0)$.
To do so, we first average $A(x,y)$
over a ring of radius $w$ and obtain,
\begin{equation}
A(w) = {\int_0^{2\pi} d\eta\,A(w\cos\eta,w\sin\eta)\over 2\pi}
= A_0 + A_2 w^2 + A_4 w^4 + \ldots
\label{eqn:aofz}
\end{equation}
where
\begin{equation}
A_0\equiv A_{00};\qquad
A_2\equiv {A_{20} + A_{22}\over 2};\qquad
A_4\equiv {3A_{40} + A_{42} + 3A_{44}\over 8}.
\label{eqn:adefs}
\end{equation}

To evaluate $A_{\rm finite}$ in terms of $A_0$, 
$A_2$, $A_4\ldots$, we must
first specify a limb-darkening law for the surface
brightness $S(w)$, which for simplicity we take to be linear,
\begin{equation}
S(w)= \biggl[1 - \Gamma
\Biggl(1 - {3\over 2}\sqrt{1 - {w^2\over\rho^2}}\Biggr)\Biggr]
{F\over \pi\rho^2},
\label{eqn:limbdarken}
\end{equation}
where $\Gamma$ is the limb-darkening coefficient and $F$ is the
source flux.
We then substitute into equation (\ref{eqn:aofz}) to obtain
\begin{equation}
A_{\rm finite} \equiv {\int_0^\rho dw\,2\pi w\,A(w)S(w)\over F}
= A_0 + {A_2\rho^2\over 2}\biggl(1 - {1\over 5}\Gamma\biggr)
+ {A_4\rho^4\over 3}\biggl(1 - {11\over 35}\Gamma\biggr) + \ldots
\label{eqn:afinite}
\end{equation}
Note that the ``limb-darkening factors'' in parentheses are
simply $(2\pi/F)\int_0^\rho dw\,S(w)w^{2n+1}$ \citep{pejcha07}, 
and therefore do not
depend on the magnification in any way.  Hence, for any given
adopted limb-darkening profile, these can be calculated just once.

Let us now assume that the field is adequately described by a 
hexadecapole.  Averaging over the four points on a $w$-ring that
are shifted by an arbitrary angle $\phi$ relative the cardinal directions,
we obtain
$$
A_{w,+} \equiv 
{1\over 4}\sum_{j=0}^3 A\biggl[w\cos\biggl(\phi+j{\pi\over 2}\biggr),
w\sin\biggl(\phi+j{\pi\over 2}\biggr)\biggr] - A_0
$$
\begin{equation}
= A_2 w^2 + {(A_{40}+A_{44})(1+\cos^2 2\phi)+ A_{42}\sin^2 2\phi\over 4}w^4,
\label{eqn:oplus}
\end{equation}
while rotating this geometry by $\pi/4$ gives,
\begin{equation}
A_{w,\times} 
= A_2 w^2 + {(A_{40}+A_{44})(1+\sin^2 2\phi)+ A_{42}\cos^2 2\phi\over 4}w^4.
\label{eqn:otimes}
\end{equation}
For a given source of size $\rho$ and position $(x_0,y_0)$,
one can therefore
determine $A_0$, $A_{\rho,+}$, $A_{\rho/2,+}$, and $A_{\rho,\times}$,
from a total of 13 point-source calculations, 
and thus derive
\begin{equation}
A_2\rho^2 = {16 A_{\rho/2,+} -A_{\rho,+}\over 3},
\qquad
A_4\rho^4 = {A_{\rho,+} + A_{\rho,\times}\over 2} - A_2\rho^2,
\label{eqn:a2a4}
\end{equation}
which can then be substituted 
into equation (\ref{eqn:afinite}) to
obtain $A_{\rm finite}(\rho,x_0,y_0)$.

{\section{Range of Validity}
\label{sec:valid}}

Clearly, this approximation cannot be used when the source 
lies on a caustic, but how close can the source be before
the approximation breaks down?  The breakdown will be driven by the leading
term of the caustic's singularity, so it is sufficient to
examine idealized cases whose leading-order behavior is the
same as that of real caustics.  More formally, one can write the 
magnification field as the linear sum of an idealized singularity and
a more complicated, but well-behaved field.   Only the former will
contribute to the breakdown of the approximation.
Since equations (\ref{eqn:afinite}), (\ref{eqn:oplus}), (\ref{eqn:otimes}), 
and (\ref{eqn:a2a4})
are strictly linear, this decomposition is absolutely rigorous.
Caustic singularities come in two basic
varieties, fold caustics and cusps.  The former diverge
as $(\Delta u_\perp)^{-1/2}$ as one approaches the caustic from
the inside, where $\Delta u_\perp$ is the perpendicular
distance to the fold.  The latter are much more complex.  They
diverge as $(\Delta u)^{-1}$ in the immediate neighborhood of the
cusp as one approaches it from the outside.  However, at greater
distances, they develop into long ``fingers'' \citep{gouldloeb92,pejcha07}.

{\subsection{Fold Caustic}
\label{sec:fold}}

For simplicity, I consider a uniform source
that lies a distance $z\rho$ (where $\rho$ is the source radius) 
from the fold.  I begin by assuming that the magnification is dominated by 
the two ``new images'' that meet on the critical curve.  I then discuss
how the result is changed when this assumption is relaxed.
The magnified flux is then given by
\begin{equation}
{A(z)\over A_0} = {2\over \pi}\int_{-1}^1 d x \sqrt{1-x^2\over 1 + x/z}=
{2\over \pi}\int_{-1}^1 d x \sqrt{1-x^2}\sum_{m=0}^\infty
{(2m-1)!!\over 2^m m!}\biggl({x\over z}\biggr)^m,
\label{eqn:fold1}
\end{equation}
which after some algebra [e.g., 
$\int_0^1 dy y^m(1-y)^n=m!n!/(m+n+1)!$], 
simplifies to
\begin{equation}
{A(z)\over A_0} =\sum_{n=0}^\infty 
{(4n-1)!!(2n-1)!!\over 2^{3n}(2n)!(n+1)!}z^{-2n},
\label{eqn:fold4}
\end{equation}
where $n = m/2$.  That is,
\begin{equation}
{A(z)\over A_0} = 1 
+ {3\over 2^5}z^{-2}
+ {35\over 2^{10}}z^{-4}
+ {1155\over 2^{16}}z^{-6}
+ {45045\over 2^{22}}z^{-8} + \ldots
\label{eqn:fold5}
\end{equation}
Hence the error due to the hexadecapole approximation is
\begin{equation}
{\Delta A(z)\over A_0} = {1155\over 2^{16}}z^{-6} = 2.8\times 10^{-4}
\biggl({z\over 2}\biggr)^{-6}.
\label{eqn:fold6}
\end{equation}

If one is ``sufficiently near'' a fold caustic, the magnification
will always be dominated by the two ``new images''.  However,
for planetary caustics, this will generally no longer be the case even at
one or two source radii from the caustic.  The net effect is to ``dilute''
the caustic divergence and so to make the the
hexadecapole approximation valid at even closer separations than
indicated by equation (\ref{eqn:fold6}).  See \S~\ref{sec:example}.

\subsection{{Cusp}
\label{sec:cusp}}

For simplicity, I analyze the case of a $(\Delta u)^{-1}$ divergence
and then discuss how the results may be expected to change for real
cusps.  In this case, we have
$$
{A(z)\over A_0} = {1\over 2\pi}\int_0^1 dx^2\int_0^{2\pi}d\theta
\biggl(1 + {2x\cos\theta\over z} + {x^2\over z^2}\biggr)^{-1/2}
$$
\begin{equation}
= {1\over 2\pi}\int_0^1 dx^2\int_0^{2\pi}d\theta
\sum_{m=0}^\infty (-1)^m{(2m-1)!!\over m! 2^m}\sum_{i=0}^m C^m_i
\biggl({2x\cos\theta\over z}\biggr)^i\biggl({x^2\over z^2}\biggr)^{m-i},
\label{eqn:point2}
\end{equation}
which after some algebra [e.g.,
$\langle (2\cos\theta)^{2k}\rangle = C^{2k}_k$] reduces to
\begin{equation}
{A(z)\over A_0} = 
\sum_{n=0}^\infty {z^{-2n} \over (n+1)}
\sum_{j=0}^{n} \biggl(-{1\over 2}\biggr)^{j+n}
{(2n+2j-1)!!\over (n-j)![j!]^2},
\label{eqn:point7}
\end{equation}
where, again, $n=m/2$.  That is,
\begin{equation}
{A(z)\over A_0} = 1 
+ {1\over 2^3}z^{-2}
+ {3\over 2^6}z^{-4}
+ {25\over 2^{10}}z^{-6}
+ {245\over 2^{14}}z^{-8} + \ldots,
\label{eqn:point8}
\end{equation}
so that the error due to the hexadecapole approximation is,
\begin{equation}
{\Delta A(z)\over A_0} = {25\over 2^{10}}z^{-6} = 3.8\times 10^{-4}
\biggl({z\over 2}\biggr)^{-6}.
\label{eqn:point9}
\end{equation}

Equation (\ref{eqn:point9}) gives a reasonable lower limit on how
closely one may approach a cusp using the hexadecapole approximation.
However,  because the cusp develops a linear, finger-like structure
at moderate distances, this approximation can fail well away
from the cusp.  Nevertheless, I find numerically that
the total duration of the failure is generally represented
reasonably well by equation (\ref{eqn:point9}).  See \S~\ref{sec:example}.

Thus, for both fold caustics and cusps, the hexadecapole approximation
can reduce the error below 0.1\%, except for intervals characterized
by $z=2.5$.  

{\section{Implementation}
\label{sec:implement}}

The main consideration when implementing this approximation is
to make certain that it is applied only in its range of validity.
This is easiest when one is probing an already-located minimum,
which is often the most time-consuming part of the investigation.
One can then simply plot the difference between the hexadecapole
approximation and the finite-source calculation for a {\it single model}
as a function of time, and so locate empirically the regions of
the former's range of validity.  If a ``safety zone'' is placed around
the finite-source regions, then there is little danger that it will
be crossed during the minor excursions that occur while probing
a minimum.  It is straightforward to determine automatically whether
such unexpected crossings are occurring simply by comparing
the finite-source and hexadecapole calculations for the first and last
points of each finite-source region.  If these do not agree, the
``safety zone'' has been crossed.

One must be more careful when applying this method to blind searches
because it is harder to determine whether any particular
stretch of the lightcurve is either crossing or nearby a caustic.  In some
cases, this will be straightforward and in others more difficult.
The one general point to note is that the same ``safety zone''
check can be made.

Finally, I remark that in some of the lightcurve regions where the
hexadecapole approximation is valid, it may be overkill: the
quadrupole or even monopole (point-source) approximations may be perfectly
satisfactory.
Again, it is straightforward to find these subregions, simply
by mapping the hexadecapole/quadrupole and hexadecapole/monopole
differences for a single model.

{\section{An Illustration of the Method}
\label{sec:example}}

\begin{figure}
\plotone{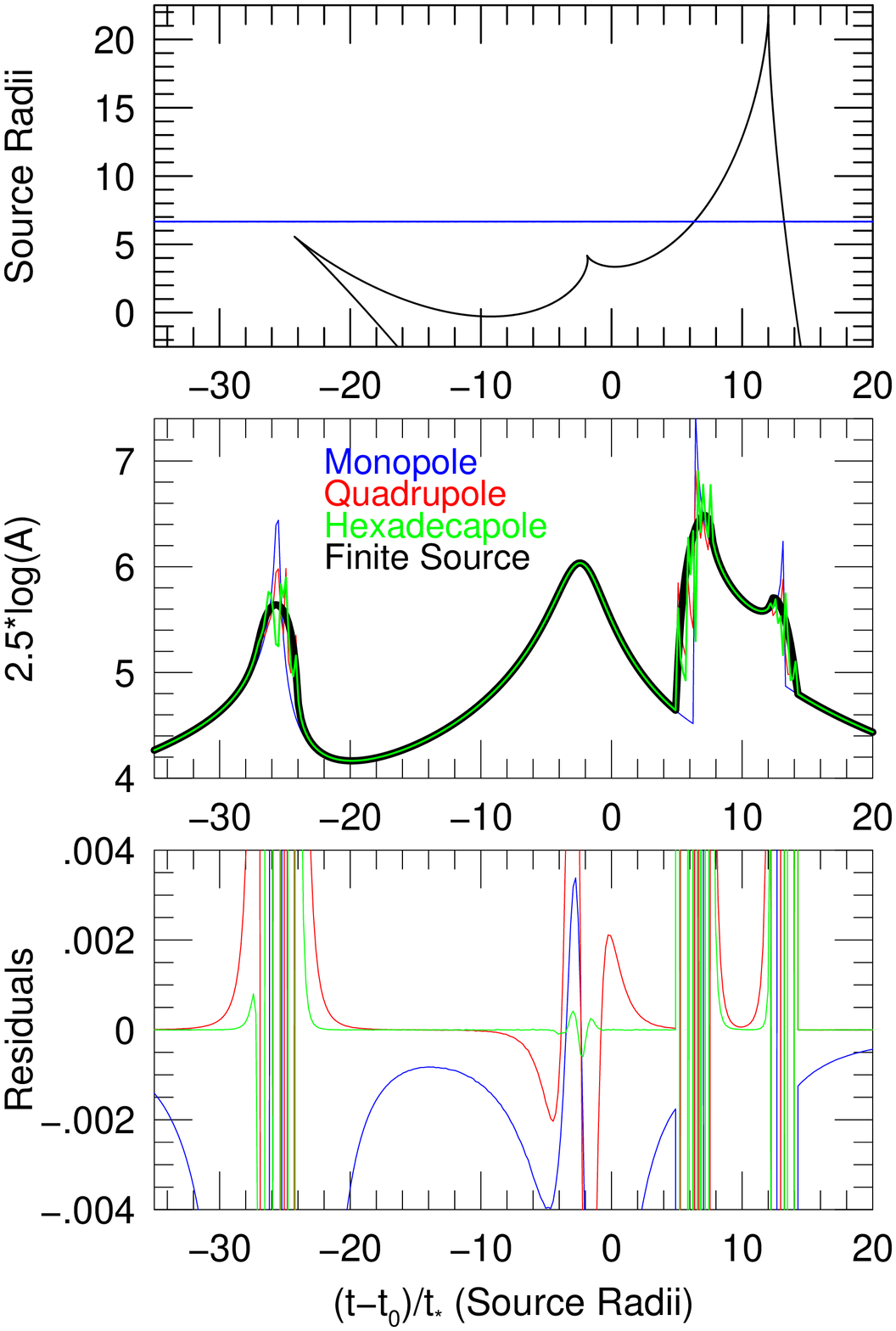}
\caption{\label{fig:ex1}
Top: source trajectory ({\it blue}) through caustic geometry ({\it black})
of a simulated high-magnification microlensing event, in units of the
source radius.  Middle: resulting lightcurve as found from
full finite-source calculation ({\it black}) and the 
monopole ({\it blue}), quadrupole ({\it red}), and hexadecapole ({\it green})
approximations.  Bottom: residuals of the three approximations relative
to the full calculation.  See \S~\ref{sec:example} for full discussion.
}
\end{figure}

\begin{figure}
\plotone{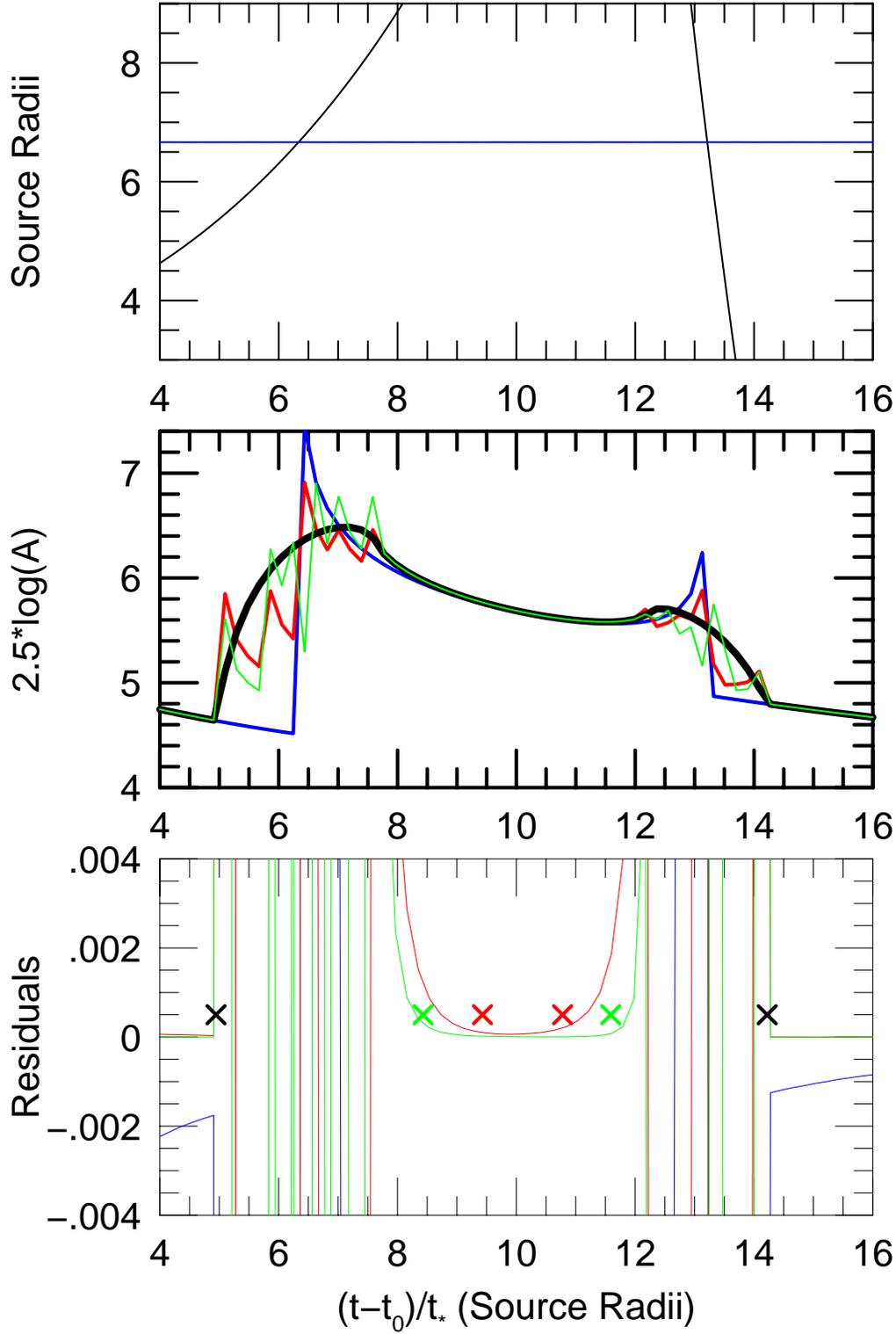}
\caption{\label{fig:ex2}
Zoom of Fig.~\ref{fig:ex1}, focusing on region of the two caustic crossings.
Crosses in middle panel show predictions for range of validity of the
quadrupole ({\it red}) and hexadecapole ({\it green}) approximations
inside the caustic and for both ({\it black}) outside.  
See \S~\ref{sec:example} for full discussion.
}
\end{figure}

Figures \ref{fig:ex1} and \ref{fig:ex2} give a practical example
of the hexadecapole approximation.  The top panel of Figure \ref{fig:ex1}
shows a caustic geometry ({\it black}) that is based loosely on
the caustic in the planetary event OGLE-2005-BLG-071 
\citep{ob05071}, but with a different
source trajectory ({\it blue}) that has been chosen the maximize the
number of illustrative ``features'':  the source passes by two cusps
and then enters and exits the caustic.

The middle panel shows the resulting lightcurve ({\it black})
together with three successive levels of approximation:
monopole (i.e., point-source {\it blue}), quadrupole ({\it red}),
and hexadecapole ({\it green}).  The bottom panel shows the
residuals of the latter three relative to the first.

There are several points to note.  First, the hexadecapole approximation
works extremely well over the entire region shown except for a
few source-radius crossing times in the immediate vicinity of the
first cusp approach and the two caustic crossings.  Second, the
point-source approximation basically does not work at all over the
entire region shown, at least if one is attempting to achieve
precisions of 0.1\% (which is typically required). Third, there
are significant regions where the quadrupole approximation is adequate.

Figure \ref{fig:ex2} is a zoom of Figure \ref{fig:ex1}, focusing on the
region of the two caustic crossings.  The crosses in the bottom
panel show the predictions for the breakdown of the
quadrupole ({\it red}) and hexadecapole ({\it green}) approximation
inside the caustic (eqs.~[\ref{eqn:fold5}] and [\ref{eqn:fold6}]), 
assuming one is
trying to achieve a precision of 0.05\%.  The {\it black} crosses
indicate the prediction for breakdown outside the caustic, namely
one source radius from the caustic.

Both the quadrupole and hexadecapole approximations prove to be
too conservative in that the approximations remain valid substantially
closer to the caustic than expected.  The reason for this is
clear from the middle panel: at the points of the predicted breakdown,
the underlying magnification profile is no longer dominated by
the square-root singularity that is produced by the two ``new
images''.  Recall from \S~\ref{sec:fold} that equations 
(\ref{eqn:fold5}) and (\ref{eqn:fold6})
were specifically derived under the assumption these images would dominate
the magnification.

Returning to Figure \ref{fig:ex1}, the situation is more complex
for the cusp approaches.  Equation (\ref{eqn:point9}) predicts that
the hexadecapole approximation will break down from $-25.8$ to $-22.7$.
The actual breakdown is from $-27.6$ to $-23.2$.  This displacement
to the left reflects the ``finger'' of enhanced magnification that
extends from the cusp axis, as indicated in the top panel.  In fact,
I find that for trajectories passing farther from this caustic,
the breakdown occurs at even earlier times, again corresponding to
the intersection of the trajectory with the cusp axis (rather
than the point of closest approach).  Moreover, the breakdown
continues to occur even when the point of closest approach is
well beyond the separation predicted by equation (\ref{eqn:point9}).
Nevertheless, the full duration of the breakdown is never much 
greater than $2z$ as predicted by this equation.  In brief, for
cusps, the hexadecapole approximation does work except for brief
intervals whose duration is given by equation (\ref{eqn:point9}), but
the time of breakdown is not accurately predicted by this equation.
Similar remarks apply to the quadrupole approximation.

{\section{Application to Lenses With Orbital Motion}
\label{sec:orbits}}

While the hexadecapole approximation can save substantial computation
time in a wide range of cases, it may be especially useful for
planetary lenses with measurable orbital motion by rendering them 
accessible to the ``magnification map'' technique \citep{ob04343}.
Magnification maps are potentially very powerful.  One first
chooses ``map parameters'', i.e., $(b,q)$ for a single planet or 
$(b_1,q_1,b_2,q_2,\phi)$ for a triple system, where $\phi$
is the angle between the two planets.  Then one shoots the entire
Einstein ring (out to a specified width corresponding to, say,
magnification $A=100$).  One then both stores the individual
rays and tiles the source plane with hexagonal pixels, keeping
track of the number of rays landing in each pixel.  Pixels landing
wholly within the source are evaluated at their centroid,
while pixels that cross the boundary are evaluated on a ray-by-ray
basis.  Using this map, one can minimize over the remaining
microlensing variables (time of closest approach $t_0$, impact parameter
$u_0$, Einstein timescale $t_{\rm E}$, source trajectory angle $\alpha$,
source size $\rho$, as well as parallax and xallarap parameters, if
these are needed).  Each map can be created in a few seconds and
fully explored in a few minutes, thereby permitting 
a rapid Markov-chain walk through map space.

The drawback is that, to date, magnification maps have not been
applicable to lenses with significant orbital motion: to the
extent that the lens separation changes during the event, different
maps would be needed at different phases of the event, potentially
a very large number of them.  In some
cases, this problem is now completely resolved.  For example,
OGLE-2005-BLG-071 exhibits some signature of
rotation \citep{dong08}, which had been difficult to probe
simultaneously with finite-source effects.  Because the source
trajectory comes no closer than $z=10$ source radii from the
cusp, this event can now be handled completely in the hexadecapole
approximation.

However, even for events with one or several caustic features that
require finite-source calculations, it will now be possible to
evaluate these with maps.  Each caustic feature lasts only about
the time required for the source to cross its own diameter, which
is typically a few hours.  The orbital motion during these features
is negligible, implying that the lens geometry 
can be adequately represented by a single map.
Several maps can be created, one for each feature occurring at
different times.  The points ``between features'' can be evaluated
in the hexadecapole approximation, which allows a continuously
evolving lens geometry.

\section{{Conclusions}
\label{sec:conclude}}

I have identified an intermediate regime between the ones where
finite-source effects are dominant and negligible, respectively.
In this regime, magnifications can be evaluated with very
high precision using a simple hexadecapole approximation, for
which I give a specific prescription.  Outside of small
(few source-diameter crossing times) regions associated with
caustic crossings and cusp approaches,
the approximation has a fractional error of well under 0.1\%.
Some events can now be analyzed without any traditional finite-source
calculations, while for others these calculations will be drastically
reduced.  In particular, by restricting the regions that do absolutely 
require finite-source calculation to a few isolated zones, this
approximation opens the possibility of applying ``magnification maps''
to planetary systems experiencing significant orbital motion.


\acknowledgments

I thank Subo Dong for seminal discussions.  The paper benefited greatly
from the suggestions of an anonymous referee.
This work was supported by NSF grant AST 042758.

\end{document}